\documentclass[reprint,aps,twocolumn, superscriptaddress]{revtex4-1}

\usepackage[hidelinks]{hyperref} 
\usepackage{graphicx}
\usepackage{amsmath}
\usepackage{amssymb}
\usepackage{siunitx}
\usepackage{braket}

\DeclareGraphicsExtensions{.png,.jpg,.eps}
\usepackage{xcolor}
\hypersetup{
    colorlinks,
    linkcolor={red!50!black},
    citecolor={blue!80!black},
    urlcolor={blue!80!black}
}

\renewcommand{\bm}{\boldsymbol}

\begin{document}

\author{Bastian Zinkl}
\affiliation{Institute for Theoretical Physics, ETH Zurich, 8093 Zurich, Switzerland}

\author{Manfred Sigrist}
\affiliation{Institute for Theoretical Physics, ETH Zurich, 8093 Zurich, Switzerland}

\title{Impurity induced double transitions for accidentally degenerate unconventional pairing states}

\date{\today}

\begin{abstract}
Non-magnetic impurities can lift the accidental degeneracy of unconventional pairing states, such as the $(d + i g)$-wave state recently proposed for Sr$_2$RuO$_4$. 
This type of effect would lead to a superconducting double transition upon impurity doping. In a model calculation it is shown how this behavior depends on material parameters 
and how it could be detected. 
\end{abstract}

\maketitle



\section*{Introduction}
The ideal proposal for the symmetry of the order parameter of an unconventional superconductor should have the ability to explain all its specific experimental signatures. 
In the case of Sr$_2$RuO$_4$, this high standard has turned out to be most challenging. Even the candidate order parameter considered as promising over a long time, the spin-triplet chiral $p$-wave state \cite{luke1998time, mackenzie2003superconductivity, xia2006high, maeno2011evaluation}, has recently been questioned by contradictory experiments indicating spin-singlet pairing \cite{pustogow2019constraints, ishida2020reduction}. This has prompted new proposals for the pairing symmetries, some of which have quickly gained prominence, such as the even-parity, spin-singlet, time reversal symmetry breaking superposition of $d_{x^2 - y^2}$ and $g_{xy(x^2 - y^2)}$, the $ (d + i g) $-wave state \cite{kivelson2020proposal, ghosh2020thermodynamic, willa2020symmetry}. In contrast to the chiral $p$-wave state, whose two constituents, the $ p_x $ and $ p_y $-component, are degenerate by symmetry, the $ (d+ig)$-wave state has to rely on an accidental degeneracy, because $d_{x^2 - y^2}$ and $g_{xy(x^2 - y^2)}$ belong to different representations of the tetragonal point group. 
In our study, we scrutinize the $ (d+ig)$-wave specifically for this aspect of degeneracy in view of disorder effects. For this purpose, we formulated a single-band tight-binding model and apply the self-consistent $T$-matrix approximation in order to take the effect of impurity scattering on the superconducting phase into account. In this way we examine the behavior of the two pairing channels, in particular, the splitting of their transition temperatures. In the case of a double transition, we also analyze the resulting specific heat signatures.

\section*{Model of a $\bm{(d+ig)}$-wave superconductor}

\subsection*{Tight-binding model}
\begin{figure}[t!]
 \centering
	\includegraphics[width=0.90\columnwidth]{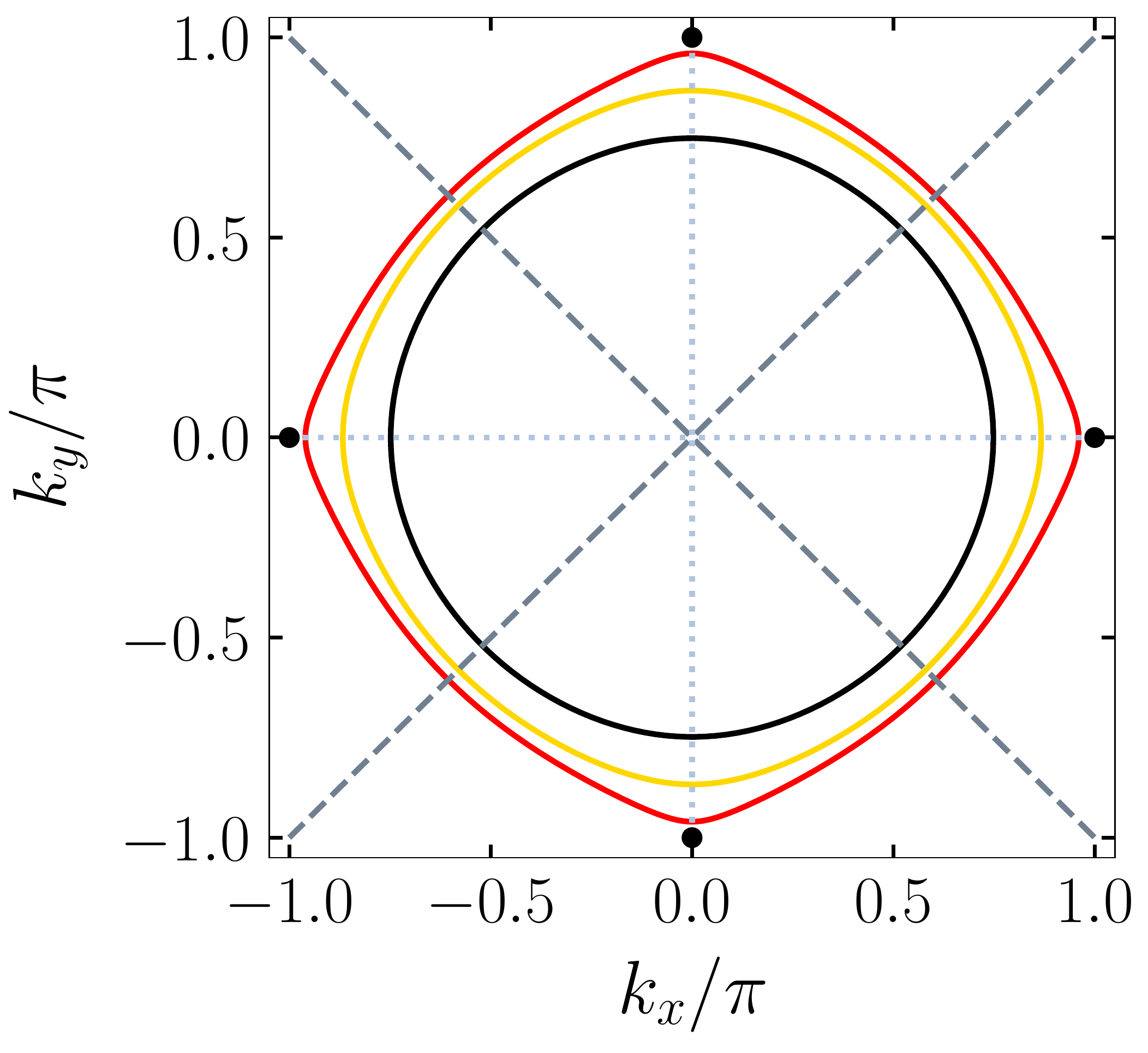}
	\caption{
		Fermi surfaces for $\mu = 0.25$ (black), $\mu = 0.925$ (gold) and $\mu = 1.175$ (red). The gap zeros of the $d$-wave are represented by the diagonal dashed lines (grey). The additional zeros of the $g$-wave are given by the horizontal and vertical dotted lines (light grey). The van-Hove points with a diverging density of states are marked by the black dots.
	}
	\label{fig:FS}
\end{figure}

We consider a single-band tight-binding model on a two-dimensional square-lattice, which includes nearest-neighbor (NN) and next-nearest-neighbor (NNN) hopping. In momentum space the Hamiltonian reads
\begin{align}
	\mathcal{H} = \sum_{\bm{k}, s} \xi_{\bm{k}} c^{\dagger}_{\bm{k}, s} c_{\bm{k}, s} + V_{\text{pair}}, 
\end{align}
where $c^{\dagger}_{\bm{k}, s}$ ($c_{\bm{k}, s}$) denotes the creation (annihilation) operator of an electron with spin $s = \uparrow, \downarrow$ and momentum $\bm{k} = (k_x, k_y)$. The dispersion, which is chosen to qualitatively resemble the genuinely two-dimensional $\gamma$ band of Sr$_2$RuO$_4$, is given by 
\begin{align}
	\xi_{\bm{k}} = -2 t (\cos k_x + \cos k_y) - 4t' \cos k_x \cos k_y - \mu, 
\end{align}
with $\mu$ as the chemical potential and hopping matrix elements $t = 1$ (unit of energy) and $t' = 0.3$ (the lattice constant $ a $ is taken to unity). In Fig.~\ref{fig:FS} we 
show the Fermi surface (FS) for varying chemical potentials. 

The pairing potential $V_{\text{pair}}$ is restricted to the spin-singlet channel, 
\begin{align}
	V_{\text{pair}} = \sum_{\substack{\bm{k}, \bm{k'} \\ s_1, s_2}} V_{\bm{k}\bm{k'}} c^{\dagger}_{\bm{k}, s_1} c^{\dagger}_{-\bm{k}, -s_1} c_{-\bm{k'}, -s_2} c_{\bm{k'}, s_2},   
\end{align}
where the orbital structure is given by $V_{\bm{k}\bm{k'}}$. With our focus on the $(d+ig)$-wave \footnote{Another even-parity state is for instance the superposition of extended $s$-wave and $d$-wave, for which we simply use the basis function $\Phi_s(\bm{k}) = \cos k_x + \cos k_y$.}, we introduce
\begin{align}
	 V_{\bm{k}\bm{k'}} = \sum_{a = d, g} V_a \Phi_a (\bm{k}) \Phi_a (\bm{k}'), 
\end{align}
where the even-parity basis functions are
\begin{align}
	 \Phi_d (\bm{k}) &= \cos k_x - \cos k_y, \\[2mm]
	 \Phi_g (\bm{k}) &= \sin k_x \sin k_y (\cos k_x - \cos k_y).
\end{align}
After the standard mean-field decoupling of the pairing potential, the minimization of the free energy leads naturally to the quasiparticle gap function 
\begin{align}
	\Delta_{\bm{k}} = &\Delta_{d} \left( \cos k_x - \cos k_y \right) \nonumber \\[2mm] 
	&\pm i \Delta_{g} \sin k_x \sin k_y \left( \cos k_x - \cos k_y \right),
\end{align}
which breaks time-reversal symmetry.
The coefficients $\Delta_{d,g}$ are obtained by solving the self-consistency equation,
\begin{align}
	\begin{pmatrix}
		\Delta_d \\ \Delta_g 
	\end{pmatrix} = \sum_{\bm{k}} \mathcal{C}_{\bm{k}} 
	\begin{pmatrix}
		V_d & 0 \\ 0 & V_g \sin^2 k_x \sin^2 k_y  
	\end{pmatrix} \begin{pmatrix}
		\Delta_d \\ \Delta_g 
	\end{pmatrix} .     \label{eqn:gapeq}
\end{align}
The factor $C_{\bm{k}}$ takes the form
\begin{align}
	\mathcal{C}_{\bm{k}} = -T \sum_n \frac{\left(\cos k_x - \cos k_y \right)^2}{\tilde{\omega}_n^2 + \xi_{\bm{k}}^2 + |\Delta_{\bm{k}}|^2}, \label{eqn:gapcpl}
\end{align}
where $T$ is the temperature and the renormalized Matsubara frequencies $\tilde{\omega}_n$ are different from the standard Fermionic ones, $\omega_n = (2n+1)\pi k_B T$, if disorder is present, as defined in Eq.~\eqref{eqn:renMats}.

\subsection*{Disorder - T-matrix approximation}
Disorder is introduced through non-magnetic impurities with a point-like potential leading exclusively to $s$-wave scattering. As we would like to explore the whole range of scattering potential strengths, meaning also the unitary limit where the potential exceeds the band width, we employ a $T$-matrix approach, which includes multiple scatterings at the same impurity. The $T$-matrix is defined by
\begin{align}
	T_{\bm{k} \bm{k}'}(i\omega_n) = U_{\bm{k} \bm{k}'} + \sum_{\bm{k}''} U_{\bm{k} \bm{k}''} G(\bm{k}'', i\omega_n) T_{\bm{k}'' \bm{k}'}(i\omega_n), 
\end{align}
where $U_{\bm{k} \bm{k}'} $ is the impurity potential in $\bm{k}$ space and $G(\bm{k}, i\omega_n)$ the (normal) electron Green's function. Note that we have omitted off-diagonal terms involving the anomalous Green's function, since they vanish for unconventional states. For $s$-wave scattering both $U_{\bm{k} \bm{k}'} $ and the $T$ matrix are scalar in momentum space, 
\begin{align}
	U_{\bm{k} \bm{k}'} = U, \quad T_{\bm{k} \bm{k}'}(i\omega_n) = T(i\omega_n).
\end{align}
We may restrict ourselves to low impurity concentrations $c$ such that we can neglect impurity interference effects, because superconductivity is rather quickly suppressed by disorder, once the mean free path becomes comparable to the zero-temperature coherence length. 
Hence, the self-energy reads 
\begin{align}
	\Sigma (i\omega_n) = c T(i\omega_n), 
\end{align}
which renormalizes the Matsubara frequencies, 
\begin{align}
	i \tilde{\omega}_n = i \omega_n - \Sigma (i\omega_n). \label{eqn:renMats}
\end{align}
Using the renormalized frequencies $\tilde{\omega}_n$ in the self-consistent gap equation [Eqs.~(\ref{eqn:gapeq}, \ref{eqn:gapcpl})] enables us to examine the influence of disorder on the superposition of unconventional pairing states.

\begin{figure}[t!]
 \centering
	\includegraphics[width=0.95\columnwidth]{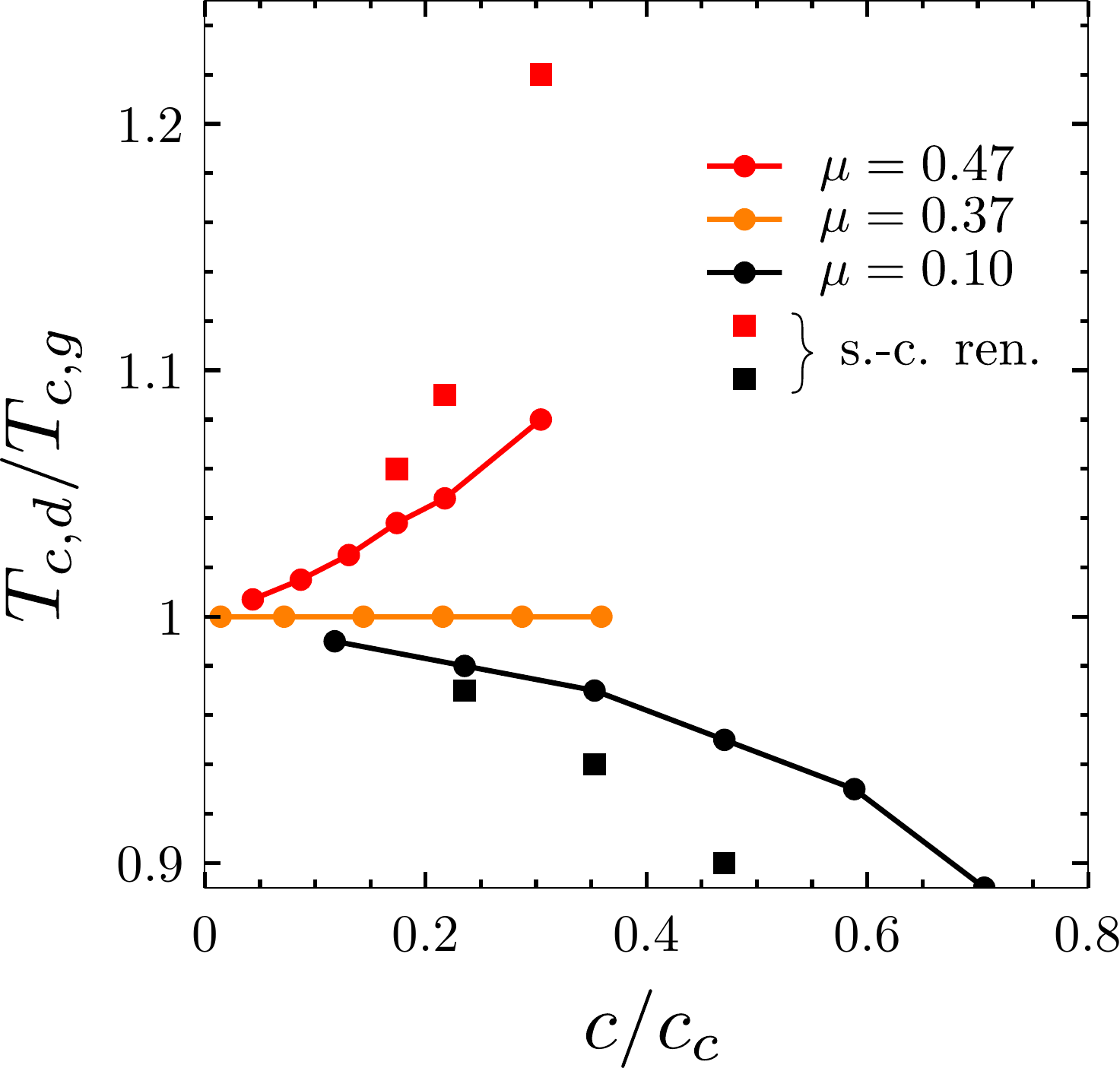}
	\caption{
		The ratio of the critical temperatures, $T_{c,d}/T_{c,g}$, as a function of the impurity concentration for different values of the chemical potential, $\mu$. We normalize the concentration values by $c_c$, which is the average of the critical concentrations, $(c_{c,d} + c_{c,g})/2$. The bare (renormalized) critical temperatures obtained from the linearized (full) self-consistent gap equations are given by the dots (squares).
	}
	\label{fig:TcImp}
\end{figure}

\subsection*{Critical temperatures $\bm{T_{c,d}}$ and $\bm{T_{c,g}}$}
For two pairing states, which belong to different representations, such as the $d$- and $g$-wave states, the respective {\it bare} critical temperatures, $T_{c,d}$ and $T_{c,g}$, are generally different. In App.~\ref{sec:app_s+id} we also discuss briefly the related case of the $(s+id)$-wave.\par
We now assume that the critical temperatures coincide in the clean system and enforce this in our model by fine-tuning the coupling strengths $ V_{d,g} $ in the pairing interaction accordingly. Focussing on the behavior of the bare critical temperatures, $ T_{c,d} $ and $ T_{c,g} $, under the influence of disorder, we solve the linearized gap equation [Eq.~\eqref{eqn:gapeq}], which decouples for the two channels. The ratio $T_{c,d} / T_{c,g}$ displayed in Fig.~\ref{fig:TcImp} (circles) reveals two regimes, if we vary the chemical potential. For $ \mu = 0.25 $ (smallest FS) the ratio $T_{c,d} / T_{c,g}$ decreases upon growing impurity concentration $ c$, while it increases for $ \mu = 1.175 $ (largest FS close to van Hove points). No change of the ratio is seen for $ \mu = 0.925 $. Thus, there is a fine-tuned FS where the ''degeneracy'' remains untouched. 

The difference in the behavior is reflected in the coherence lengths of the two pairing states, which depend on the position of the FS. A simple estimate of the zero-temperature coherence length $ \xi $ for a given gap function can be obtained from 
\begin{align}
	\xi^2 = \frac{\sum_{\bm{k}} |\nabla_k \frac{\Delta_{\bm{k}}}{E_{\bm{k}}}|^2  }{\sum_{\bm{k}}|\frac{\Delta_{\bm{k}}}{E_{\bm{k}}}|^2} .
\end{align}
For larger coherence lengths $ T_c $ suffers faster suppression with increasing $ c$. Consistently, we find $ \xi_d/\xi_g \approx 1.09 $ for $ \mu = 0.25 $ and $ \xi_d/\xi_g \approx 0.96 $ for $ \mu = 1.175 $. Intuitively it is clear for the latter case that the $d$-wave state can profit from the enlarged density of states at the van Hove points (small Fermi velocity), while the $ g$-wave state has nodes there. Hence, the $d$-wave state is more tightly bound. However, on more genuine Fermi surfaces pairing states of higher angular momentum have in general shorter coherence lengths for a given critical temperature. 

The splitting of the bare critical temperatures implies the occurrence of two consecutive superconducting transitions: First, into the superconducting phase and then breaking time reversal symmetry. The second transition, however, does not happen at the lower of the two bare $ T_c $, but at a renormalized critical temperature, because the second order parameter has to nucleate in the presence of the first one. Thus, to determine the real onset of the second order parameter we have to solve the full self-consistency equation [Eq.~\eqref{eqn:gapeq}] for $\Delta_d$ and $\Delta_g$. The renormalization of the critical temperatures, indicated by squares in Fig.~\ref{fig:TcImp}, yields a larger splitting of the two transitions than the ratio $T_{c,d} / T_{c,g}$ would suggest. Due to the presence of the first order parameter large parts of the states at the FS are consumed leaving a strongly reduced density of low-energy states available for the second order parameter. 

\subsection*{Specific heat for the double transition}
There are few ways of observing superconducting double transitions. Traditionally, specific heat has been a hallmark of such a feature in many unconventional superconductors. Thus, we would like to show here that the impurity induced split of the transition could leave an observable signature in the specific heat. We use our Green's function formalism and linear response theory \cite{luttinger1960ground, keller1988free}, as shown in App.~\ref{sec:AppCV}.\par

\begin{figure}[t!]
 \centering
	\includegraphics[width=0.9\columnwidth]{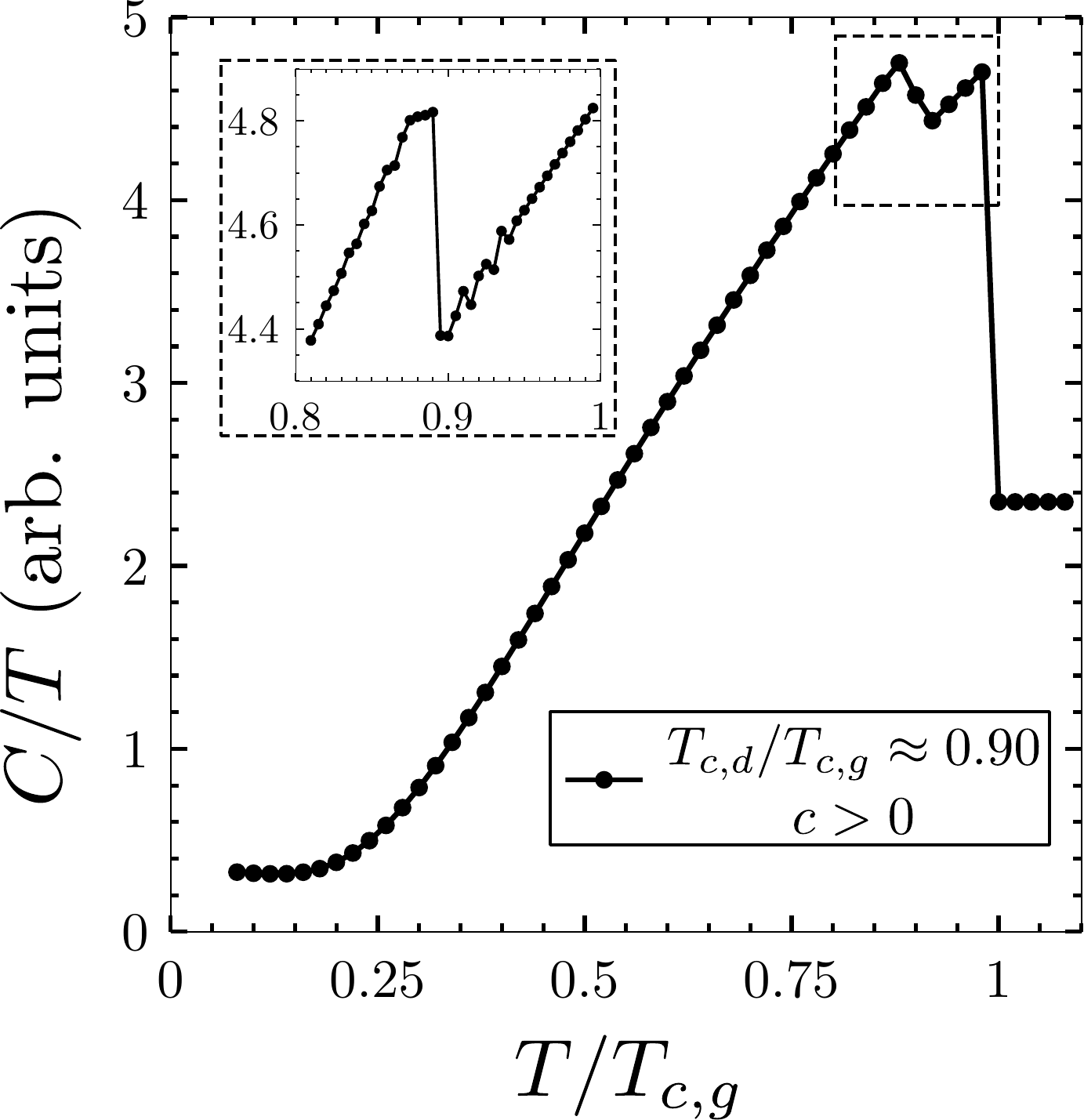}
	\caption{
		The specific heat divided through temperature, $C/T$, as a function of temperature $T$ for a finite impurity concentration, $c > 0$. The ratio of the critical temperatures is given by $T_{c,d}/T_{c,g} \approx 0.90$ and the chemical potential by $\mu = 0.10$. The inset (dashed lines) zooms in at the second jump of the specific heat, where the $d$-wave nucleates. 
	}
	\label{fig:CvJump}
\end{figure}

We consider here the situation $ T_{c,g} > T_{c,d}$, where the first transition leads to a $ g $-wave phase and the second to the time reversal symmetry breaking $(d+ig)$-wave phase. Fig.~\ref{fig:CvJump} depicts the temperature dependence of the specific heat, $C/T$. Clearly a second anomaly is visible below the onset of superconductivity (see also inset). In our calculation the second jump is roughly 20\% of the first one and both transitions are of second order. 
Furthermore, $ C/T $ reaches a finite value in the zero-temperature limit due to the finite zero-energy density of states induced by the disorder. 

\subsection*{Conclusion}
Our work highlights how non-magnetic disorder influences the transition temperatures of accidentally or nearly degenerate unconventional pairing channels. Generally, the two pairing states would show a different suppression of their critical temperatures under disorder, which would yield in turn a superconducting double transition. Such a double transition would be visible in the specific heat, as shown in Fig.~\ref{fig:CvJump}. However, since time reversal symmetry breaking would only occur at the second transition, $\mu$SR zero-field relaxation and polar Kerr effect measurements would be optimal tools to detect whether the appearance of intrinsic magnetic properties separates from the onset of superconductivity. Similarly, the renormalization of the ultrasound velocity of transverse modes would be a way to see the second transition. 
So far no such features have been reported and should therefore be indeed a target of measurements. 
The scenario based on the $(d+ig)$-wave phase for Sr$_2$RuO$_4$ relies on fine-tuning in the clean limit. To keep the degeneracy under disorder would mean to impose a second fine-tuning constraint. 

\section*{Acknowledgements}
We would like to thank Mark H. Fischer and Roland Willa for many useful discussions. 
This work was financially supported by the Swiss National Science Foundation (SNSF) through Division II (No. 184739).

\appendix
\section{Disorder effect on the ($\bm{s+id}$)-wave phase} \label{sec:app_s+id}

For completeness we address here also an alternative state proposed, which constitutes  
the superposition of the extended $s$-wave and $d$-wave states which would not be degenerate by symmetry. The gap equations read 
\begin{align}
	\begin{pmatrix}
		\Delta_s \\ \Delta_d 
	\end{pmatrix} = \sum_{\bm{k}} \mathcal{C}'_{\bm{k}} 
	\begin{pmatrix}
		V_s \Phi_s(\bm{k})^2 & 0 \\ 0 & V_d \Phi_d(\bm{k})^2
	\end{pmatrix} \begin{pmatrix}
		\Delta_s \\ \Delta_d 
	\end{pmatrix}, 
\end{align}
with
\begin{align}
	\Phi_s(\bm{k}) &= \cos k_x + \cos k_y, \\[2mm]
	\Phi_d(\bm{k}) &= \cos k_x - \cos k_y, 
\end{align}
and 
\begin{align}
	\mathcal{C}'_{\bm{k}} &= -T \sum_n \frac{1}{\tilde{\omega}_n^2 + \xi_{\bm{k}}^2 + |\Delta_{\bm{k}}|^2}.
\end{align}
As explained in the main text, we calculate the bare critical temperatures, $T_{c,s}$ and $T_{c,d}$, from the decoupled linearized gap equations of the two pairing channels. As a representative case we chose $\mu = 1.175$ (largest FS in Fig.~\ref{fig:FS}) and list the results for different impurity concentrations in Table \ref{tab:Tcs_Tcd}.
 \begin{table}[b!]
	\begin{tabular}{| c |  c |}
	 \hline 
  	   \rule{0pt}{4mm} $c/c_c$  & $T_{c,s}/T_{c,d}$ \\ [1mm]
 	 \hline 
 	  \rule{0pt}{4mm}0.11  & 0.887 \\ [1mm]
 	  0.16  & 0.824 \\ [1mm]
 	  0.22  & 0.755 \\ [1mm]
     \hline
	\end{tabular}
	  \caption{The ratio of critical temperatures of the $(s+id)$-wave at $\mu=1.175$ as a function of the impurity concentration $c$. The average of the critical concentrations, $c_{c,s} \approx 0.057$ and $c_{c,d} \approx 0.128$, is denoted by $c_c$.
		} \label{tab:Tcs_Tcd}
\end{table} 
The impurity concentration is normalized by the averaged critical concentration $c_c = (c_s + c_d)/2$. Assuming degeneracy in the clean system, we find that the ratio $T_{c,s}/T_{c,d}$ is decreasing as a function of the impurity concentration, which is in line with the ratio of coherence lengths, $\xi_s /\xi_d \approx 1.36$. We checked that the behavior of $T_{c,s}/T_{c,d}$ to decrease under impurity doping is independent of $\mu$, in contrast to the ($d+ig$) pairing state [cf. Fig.~\ref{fig:TcImp}]. 

\section{Calculation of the specific heat in disordered systems} \label{sec:AppCV}
For the derivation of the specific heat we employ the Green's function formalism and linear response theory \cite{mineev1999introduction, nomura2005theory}. 
We start with the generalized formula for the ground-state energy of an interacting electron system by Luttinger and Ward \cite{luttinger1960ground,keller1988free}. The grand potential can be written as 
\begin{align}
	\displaystyle \Omega_s = -T \sum_n \sum_{\mathbf{k}} &\Big\{ \log{(\tilde{\omega}_n^2 + \xi_{\mathbf{k}}^2 + |\Delta_{\mathbf{k}}|^2)}
	+ \Delta_{\mathbf{k}} F^{\dagger}(\mathbf{k}, i\omega_n)  \nonumber \\[1mm] 
	&+ \Sigma(i\omega_n) G(\mathbf{k}, i\omega_n) \Big\} + \Omega',
\end{align}
with $i\tilde{\omega}_n = i\omega_n - \Sigma(i\omega_n)$ and  
\begin{align}
	\displaystyle \Omega' = T \sum_{\nu} \sum_n \sum_{\mathbf{k}} \frac{1}{\nu} \Sigma_{\nu} (i \omega_n) G(\mathbf{k}, i\omega_n),
\end{align}
where $\Sigma(i\omega_n) = c T(i\omega_n) = \sum_{\nu} \Sigma_{\nu} (i \omega_n)$. 
By considering the difference between superconducting and normal state, $\Omega_s - \Omega_n$, we ensure that the sum over $n$ converges.
After calculating the self-energy self-consistently it is straightforward to determine the specific heat difference through
\begin{align}
	\frac{C_s - C_n}{T} = -\frac{\partial^2 (\Omega_s - \Omega_n)}{\partial T^2} . \label{eqn: cv}
\end{align}
The derivatives for the displayed results in Fig.~\ref{fig:CvJump} have been taken numerically. 

\bibliographystyle{apsrev4-1}
\bibliography{references}

\end{document}